\documentclass[12pt]{article}
\usepackage{cite,xcolor}
\usepackage{dcolumn}% Align table columns on decimal point
\usepackage{jheppub} % for details on the use of the package, 
\usepackage{amsmath, braket, float, slashed}
\usepackage{graphicx} % Allows including images
\usepackage{mathtools, breqn, autobreak}
\allowdisplaybreaks[1]

\newcommand{\be}{\begin{equation}}
\newcommand{\ee}{\end{equation}}
\newcommand{\bea}{\begin{eqnarray}}
\newcommand{\eea}{\end{eqnarray}}
\newcommand{\bear}{\begin{eqnarray}}
\newcommand{\eear}{\end{eqnarray}}
\newcommand{\ba}{\begin{array}}
\newcommand{\ea}{\end{array}}

\newcommand{\VQ}{Q}
\renewcommand{\arraystretch}{1.3}
\title{\vskip-3cm{\baselineskip14pt
\centerline{\hfill\rm\normalsize Fermilab-Pub-22-790-T}  
\vskip0.2cm
\centerline{\hfill\rm\normalsize 
November 3, 2022}
}
\vskip1.5cm
\boldmath \large Proton decay from quark and lepton compositeness}

\author{Beno\^it Assi and Bogdan A. Dobrescu}

\affiliation{Particle Theory Department, Fermilab, Batavia, IL 60510, USA}

\bigskip

\abstract{Within a chiral $SU(15)$ gauge theory in which the Standard Model fermions are bound states of massless preons, we show that proton-decay operators are likely induced at the compositeness scale, $\Lambda_{\rm pre}$. Our estimate of the limit imposed by searches for proton decays is $\Lambda_{\rm pre} > 10^4 \, {\rm TeV}^{1/2} \, C_8^{1/4} \, m_{\cal Q}^{1/2}$, where $C_8$ is a rescaled coefficient of an 8-prebaryon operator induced by $SU(15)$ dynamics,  and $m_{\cal Q}$ is the mass of a composite vectorlike quark. The latter has a lower limit related to the mass of a composite vectorlike lepton, which in turn is required by LHC searches to be above 1 TeV. For $C_8$ in the $10^{-5} - 1$ range,  the lower limit on $\Lambda_{\rm pre}$ varies between $3 \times 10^3$ TeV and $5 \times 10^4$ TeV. We point out that exotic proton decay modes, into a $\pi^+$ and a heavy right-handed neutrino, could be observed using the Super-Kamiokande or DUNE detectors. 
}
\begin{document} 
\maketitle

\flushbottom

%%%%%%%%%%%%%%%%%%%%%%%%%%%%%%%%%%%%%%%%%%%%%%%%%%%%%%%%%%
\section{Introduction}
\label{sec:intro}

A key question in high-energy physics is whether the observed quarks and leptons are elementary or composite particles.
Within the last three decades little progress has been made towards answering this important question.
One reason is that experimental searches for quark or lepton substructure have established that the compositeness scale must be much larger than the masses of the Standard Model (SM) fermions. The only known class of theories that may achieve that large mass 
hierarchy is based on strongly-coupled chiral gauge dynamics.  
Currently, the tools for analyzing such theories remain limited to some consistency checks, such as the 't Hooft 
anomaly matching \cite{tHooft:1979rat}, or large-$N$ arguments \cite{tHooft:1973alw,Witten:1979kh}.

Another reason for the limited progress in studying quark and lepton compositeness 
is that it has been difficult even to imagine \cite{Peccei:1985me} how the observed patterns of SM fermion masses could arise from 
a strongly-coupled chiral gauge interaction acting on some elementary fermions, historically called preons. 
The chiral nature of the theory implies that the preons are massless, and there is no obvious source for the mass splittings
between the light composite fermions, which are referred to as prebaryons and need to be identified as the quarks and leptons. 
For example, the consistency checks mentioned above indicate that 
$SU(N)$ gauge theories with a preon in the symmetric tensor representation and $N+4$ preons 
in the anti-fundamental representation produce massless prebaryons   
\cite{Dimopoulos:1980hn, Eichten:1985fs, Geng:1986xh}.

Recently, it has been proposed that the SM quarks and leptons may be composed of chiral preons transforming under a confining $SU(15)_{\rm p} $ gauge group (the index p refers to preons), and that composite Higgs fields arise as di-prebaryon bound states \cite{Dobrescu:2021fny}. 
Furthermore, the effective couplings between prebaryons and composite Higgs doublets lead to  
quark and lepton masses that are potentially compatible with the observed ones. 
This model of quark and lepton compositeness includes a preon in the conjugate symmetric tensor representation of $SU(15)_{\rm p}$, whose gauge anomaly is cancelled by 19 preons in the fundamental representation, with 15 of them also carrying the $SU(3)_c \times SU(2)_W \times U(1)_Y$ gauge charges of a SM generation. The prebaryons form exactly 3 generations of SM fermions, and several composite vectorlike fermions.

In this paper we analyze baryon-number violating effects that are likely to be produced by the dynamics responsible for  compositeness. We identify several 8-prebaryon operators that may be induced by $SU(15)_{\rm p}$ dynamics with large enough coefficients to lead to measurable proton decays. 
The ensuing constraint on the scale of quark and lepton compositeness, $\Lambda_{\rm pre}$, depends on the VEVs of some composite scalars. The latter give masses to certain composite vectorlike quarks (``Vquarks"), which in turn are related to the masses of composite vectorlike leptons (``Vleptons").  We will show that the LHC limits on Vleptons in conjunction with the proton lifetime limits can be used to derive a lower bound on $\Lambda_{\rm pre}$ roughly in the range $10^3 - 10^5$ TeV, depending on the unknown coefficients of the 8-prebaryon operators.
Under certain assumptions about the VEVs of composite scalars, we find that the spectrum of Vquarks includes two down-type ones of mass near 2 TeV, five other color-triplets of masses near an intermediate scale of about 30 TeV, and two heavier Vquarks belonging to higher $SU(3)_c$ representations.
We compute the running of the QCD coupling above the weak scale and find that it remains perturbative up to $\Lambda_{\rm pre}$, despite losing asymptotic freedom above the  intermediate scale.

The hierarchy between $\Lambda_{\rm pre}$ and the weak scale requires fine-tuning of the 
preonic ``nuclear" (referred below as prenuclear) interaction, which controls the squared masses of the di-prebaryons that  break the electroweak symmetry.
Although this fine-tuning is worrisome, it is worth recalling that similar hierarchies of scales appear in QCD. For example, 
the binding energy of the deuteron is 2.2 MeV, and the scalar neutron-proton system is unbound by an energy of $6.6 \times 10^{-2} $ MeV, {\it i.e.}, about four orders of magnitude smaller than the QCD scale \cite{Flambaum:2007mj}. 
Prenuclear interactions are surely different than usual nuclear interactions, but future progress in understanding chiral gauge theories could in principle lead to an explanation of why the weak scale is much smaller than  $\Lambda_{\rm pre}$. 
One possibility of dynamical self-tuning of the critical coupling has been explored recently \cite{Hill:2022gwl} 
in systems akin to the di-prebaryons.

Besides proton decay modes that have been searched for, such as $p \to \pi^0 e^+$, the model
of composite quarks and leptons analyzed here predicts exotic proton decay modes involving a $\pi^+$ and a heavy right-handed neutrino, which may be detector stable or may decay within the detector but  with a large displacement. These exotic decays could be searched for using existing Super-Kamiokande data, or future detectors such as DUNE. 

In Section \ref{sec:model} we present the salient features of the $SU(15)_{\rm p}$ preonic model. In Section \ref{sec:below} we estimate the mass spectrum of the Vquarks and Vleptons. The proton-decay operators and their effects are studied in Section~\ref{sec:decay}, followed by the derivations of the limit on $\Lambda_{\rm pre}$ and the computation of the running of the QCD coupling in Section~\ref{sec:perturbativity}.  The exotic proton-decay modes are proposed and analyzed in Section~\ref{sec:novel}. Our conclusions can be found in Section~\ref{sec:conc}.

%%%%%%%%%%%%%%%%%%%%%%%%%%%%%%%%%%%%%%%%%%%%%%%%%%%%%%%%%%
\section{Confining $SU(15)_{\rm p} $ gauge theory}
\label{sec:model}\setcounter{equation}{0}

The gauge symmetry of the model studied here is $SU(15)_{\rm p} \times SU(3)_c\times SU(2)_W \times U(1)_Y$,
where $SU(15)_{\rm p}$ is responsible for  preon confinenment. 
$SO(10)$ is a global symmetry that embeds the SM gauge group, $SU(3)_c\times SU(2)_W \times U(1)_Y$. 
The field content of the model is shown in Table~\ref{table:preons}.
The $\beta$ function of  $SU(15)_{\rm p}$ has a 1-loop coefficient given by
\be
b_{15} = 55 % \frac{11}{3} 15 
- \frac{2}{3} \left( 21 \, T_2(15)+  T_2(\Omega) \rule{0mm}{3.9mm}   \right) 
- \frac{1}{3} \,  T_2( {\cal A} )  =  \frac{241}{6}  ~~,
\ee
where the second Dynkin index for each
$SU(15)$ representation included here 
is $T_2(15) = 1/2$, $T_2(\Omega) =17/2$, $T_2( {\cal A} ) = 13/2$ \cite{Eichten:1982pn}.
%- 55 + 89/6 = - 241/6 
As $b_{15} > 0$, the $SU(15)_{\rm p}$ gauge interaction is asymptotically free, and becomes strongly coupled at a scale $\Lambda_{\rm pre}$.

The model includes only five preons: $\Psi, \psi_{2,3,4}, \Omega$. These are massless 2-component spin-1/2 fields,
and are the only elementary fermions lighter than the scale $\Lambda_{\rm pre}$.
One of the preons, $\Omega$, transforms in the conjugate symmetric 2-tensor representation of 
$SU(15)_{\rm p}$, whose dimension is 120. The other preons displayed in Table~\ref{table:preons},
$\psi_2$, $\psi_3$, $\psi_4$ and $\Psi$ transform in the fundamental representation of $SU(15)_{\rm p}$. 
Thus, this model belongs to the class of $SU(N)$ gauge theories with one fermion in the symmetric representation and $N+4$ fermions in the fundamental representation, which 
are expected \cite{Dimopoulos:1980hn, Eichten:1985fs} to confine while preserving the $U(N+4)$ chiral symmetry. 
Due to $SU(15)_{\rm p}$ interactions, chiral prebaryons form as bound states of three preons:
\be
\Psi\psi_i\Omega  \, , \,\, 
\psi_i\psi_j\Omega \, , \,\, 
\Psi\Psi\Omega \, , \,\, 
\label{eq:10preb}
\ee
for $i,j = 2,3,4$. 
This conclusion is not modified by the presence of the complex scalar $\cal A$, which transforms in the conjugate antisymmetric 2-tensor representation
of $SU(15)_{\rm p}$, even when its mass is comparable to $\Lambda_{\rm pre}$.
As discussed in Section \ref{sec:below}, these prebaryons include 3 generations of SM quarks and leptons, and several heavy fermions which are vectorlike under the SM gauge group.

The scalar $\cal A$ has flavor-dependent couplings to the $\psi_i$ preons:
\be
 \sum_{i\ge j=2}^{4} \lambda_{ij} \, {\cal A} \, \psi_i  \psi_j + {\rm H.c.}
\label{eq:Apsipsi}
\ee
These Yukawa couplings are responsible for the flavor dependence of the SM fermion masses. 
Upon a global $U(2)$ transformation acting on $\psi_2$ and $\psi_3$, the complex dimensionless parameters $\lambda_{ij} $ can be chosen such that $\lambda_{42} = 0$.

%%%%%%%%
\begin{table}[t]
\begin{center}
\renewcommand{\arraystretch}{1.5}
\begin{tabular}{c|c|c|c|l}\hline  
 field   &   spin  &  $SU(15)_{\rm p}$   &   $SO(10)_{\rm global}  $   &  \hspace*{0.6cm} comments
\\ \hline \hline
    $ \Psi $   &  1/2  &    15   &   $16$    &     
 \\    \cline{1-4}   
 \\   [-1.3cm]  
    $ \psi_2,  \psi_3 ,   \psi_4     $  &  1/2  &   15   &  $1 $    &  
  $\left. \rule{0mm}{11mm} \right\} $  massless  preons 
    \\   [-0.5cm]  \cline{1-4} 
     $\Omega$   &  1/2  &    $\overline{120}$   &  $1 $     &     
   \\ \hline  \hline  
 $\cal A$   &  0  &    $\overline{105}$    &  $1 $    &   flavor-dependent couplings
\\   \hline 
\end{tabular}
% \vspace*{-0.1cm}
\caption{Fields charged under the 
$SU(15)_{\rm p}$ gauge group. The global $SO(10)$ symmetry embeds the SM gauge group. 
 \\[-0.7cm] }
\label{table:preons}
\end{center}
\end{table}
%%%%%%%%%%%

The $\Psi$ fermion, which belongs to the spinorial representation of $SO(10)$, decomposes
into $SU(3)_c\times SU(2)_W \times U(1)_Y$ representations as follows:
\bear
\Psi  & = & \psi_{_U} ( \overline 3,1,-2/3) + \psi_{_Q} (3,2,+1/6) + \psi_{_E} (1,1,+1) 
\nonumber \\ [2mm]
&& + \,  \psi_{\! _D} ( \overline 3,1,+1/3) + \psi_{_L} (1,2,-1/2) +  \psi_{_1} (1,1,0) ~~~.
\label{eq:Psi}
\eear
As $\Psi$ also transforms in the fundamental representation of the unbroken $SU(15)_{\rm p}$ gauge symmetry, each of its components 
($\psi_{\! _D}, \psi_{_L}, \psi_{_U}, \psi_{_Q}, \psi_{_E}, \psi_{_1}$) belongs to the 15 of $SU(15)_{\rm p}$.
An alternative notation for $\psi_1$ can be $\psi_{_N}$, because it is a SM-singlet left-handed fermion, {\it i.e.,} a preon 
which  has the quantum numbers (other than transforming under 
$SU(15)_{\rm p}$) of  the conjugate of a ``right-handed neutrino".
The $\psi_1$ notation, however, is better suited for keeping track of the SM generations: as explained in the next Section, the
quarks and leptons of the $i$th generation are given by the prebaryons $\psi_{_Q} \psi_i \, \Omega$, 
$\psi_{_L} \psi_i \, \Omega$, etc., for $ i = 1,2,3$.

%%%%%%%%%%%%%%%%%%%%%%%%%%%%%%%%%%%%%%%%%%%%%%%%%%%%%%%%%
%%%%%%%%%%%%%%%%%%%%%%%%%%%%%%%%%%%%%%
\section{Effective theory below the compositeness scale}
\label{sec:below}\setcounter{equation}{0}

At energy scales smaller than the compositeness scale, $\Lambda_{\rm pre}$, the only fields present are 
SM gauge bosons and the $SU(15)_{\rm p}$-singlet bound states protected by the chiral symmetry. 
The latter include the chiral prebaryons  formed of three preons listed in Eq.~(\ref{eq:10preb}), as well as bound states of two or more chiral prebaryons.

Chiral prebaryons  of the type $\Psi \psi_i \Omega$, with $i = 2,3,4$, belong to the 16 representation of the global $SO(10)$ symmetry, 
so their field content under the $SU(3)_c \times SU(2)_W \times U(1)_Y$ gauge group is
\bear
\Psi \, \psi_i \, \Omega &=& \Omega_{D i} ( \overline 3,1,+1/3) + \Omega_{L i} (1,2,-1/2) +
\Omega_{U  i} ( \overline 3,1,-2/3) 
\nonumber
\\ [2mm]
&& + \, \Omega_{Q  i}  (3,2,+1/6) + \Omega_{E  i} (1,1,+1) 
+ \Omega_{1  i} (1,1,0) ~.
\eear
Here $\Omega_{D i}$, for example, is short-hand notation for the $\psi_{\! _D} \psi_{\! _i} \, \Omega$ composite fermion, where $\psi_{\! _D}$ is defined in Eq.~(\ref{eq:Psi}).
Chiral prebaryons  of the type $\Psi \Psi \Omega$ are antisymmetric under the interchange of the $SO(10)$ indices of the $\Psi\Psi$ bilinear,
so they belong to the $(16\times 16)_{\rm antisym.} = 120$ representation of $SO(10)$. 
Their decomposition in SM multiplets is the following \cite{Slansky:1981yr}:
\bear
\Psi \Psi \Omega &=& 2 \! \times \! (\overline 3,1,+1/3)    +  2 \! \times \! (1,2,-1/2)      
+ ( \overline  3,1, -2/3)  + ( 3,2, +1/6)   +  (1,1, +1)   
\nonumber \\ [2mm]  && + \, 
 (3,1,-4/3)  +  (3,2,+7/6) + (3,3,-1/3) + (6, 1,+1/3) + (8,2,+1/2) 
\nonumber \\  [2mm]    && + \;  {\rm conjugates}  ~.
\label{eq:decom}
\eear
Each of the above SM representations refers to a certain prebaryon. For example, there are two $(\overline 3,1,+1/3)$
 states,    $\Omega_{D1} $ and $\Omega_{UE} $, while their partners transforming in the conjugate 
representation  $(3,1,-1/3)$  are  $\Omega_{UD}^{(3, 1)} $  and $\Omega_{QL}^{(3, 1)} $.
Here the upper indices denote the $SU(3)_c \times SU(2)_W$ representation of the prebaryon.
To simplify the notation, prebaryons such as $\Omega_{UE } \equiv \psi_{\! _U} \psi_{\! _E} \, \Omega$, 
with  SM representations uniquely determined by the preon ones, do not have upper indices.
Note that the 120 representation of $SO(10)$ is real, which explains the presence of the conjugate representations
in the decomposition (\ref{eq:decom}). The fact that both the antisymmetric $SO(10)$ tensor and the symmetric $SU(15)$ tensor have the same dimension, 120, is coincidental and of little relevance in what follows.

Finally, the chiral prebaryons  of the type $\psi_i \psi_j \Omega$, with $i , j= 2,3,4$,  are $SO(10)$  singlets. As these are antisymmetric under $i \leftrightarrow j$, there are three of them, labelled $\Omega_{32}$, $\Omega_{42}$ and $\Omega_{43}$. Altogether, the chiral prebaryons form 3 generations of SM quarks and leptons, 12 Dirac fermions which are vectorlike under the SM gauge group, and 6 gauge-singlet Weyl fermions. 

Scalars lighter than  $\Lambda_{\rm pre}$ arise as bound states of two chiral prebaryons \cite{Dobrescu:2021fny}. 
The binding of these ``di-prebaryons" is due to $SU(15)$ remnant interactions (somewhat similar to nuclear interactions in QCD)
with additional contributions from SM gauge interactions and scalar $\cal A$ exchange.
Since all these interactions are nonconfining, the low-energy effective theory includes large Yukawa couplings of the 
di-prebaryons to its constituents. For example, the Yukawa couplings of the vectorlike fermions of SM charges $(8,2,+1/2)$, $(6, 1,+1/3)$,  $(3,2,+7/6)$  and  $(3,2,+1/6)$ are given by
\be
- y_{_{88}}    \phi_{\! _{88}}^*  \Omega_{QU }^{ (8, 2) }   \Omega_{QD }^{ (8, 2) }
- y_{_{6\bar 6}}  \phi_{\! _{6\bar 6}}^*   \Omega_{QQ }^{ (6, 1) }   \Omega_{UD }^{ (\overline  6, 1) } 
- y_{_{7/6}}   \phi_{\! _{7/6}}^*  \Omega_{QE } \Omega_{UL }  
- y_{_{\cal Q}}   \phi_{\! _{\cal Q}}^*   \Omega_{Q4 }   \Omega_{DL }     \, .
\label{eq:largeYuk}
\ee
Both $\Omega_{QU }^{ (8, 2) }$ and $\Omega_{QU }^{ (1, 2) }$ are $\psi_{\! _Q} \psi_{\! _U} \Omega$ 
weak-doublet prebaryons, but under $SU(3)_c$ one is an octet and the other one is a singlet. The same is true for the 
$\Omega_{QD}$ prebaryons.
The $ \phi_{_{88}}$ scalar is a gauge-singlet bound state of $\Omega_{QU }^{ (8, 2) }$ and $\Omega_{QD }^{ (8, 2) }$.
Similarly,   $\phi_{_{6\bar 6}}$, $\phi_{_{7/6}}$ and $\phi_{\! _{\cal Q}}$ are gauge-singlet composite scalars, and their constituents are the prebaryons that participate in the Yukawa couplings (\ref{eq:largeYuk}).
We use the notation $\phi_{\! _{\cal Q}}$ instead of $\phi_{_{1/6}}$ to emphasize that the constituents of this scalar have the same charges as the SM quark doublets.
All dimensionless parameters  $y_{_{88}}$, $y_{_{6\bar 6}}$, $y_{_{7/6}}$, $y_{_{\cal Q}}$ are estimated to be given by 
\be
y_\phi \approx \frac{4\pi}{ \sqrt{ d_\phi \ln\left( \Lambda_{\rm pre}^2/\langle \phi \rangle^2 \right) } }   ~~
\label{eq:y0}
\ee
at the scale  of the corresponding VEV $\langle \phi \rangle$ \cite{Bardeen:1989ds},
but their RGEs are different due to the different SM gauge charges of the prebaryons.
The Yukawa coupling of a generic di-prebaryon $\phi$, given   in Eq.~(\ref{eq:y0}), 
depends on the dimensionality $d_\phi$ of the $SU(3)_c \times SU(2)_W$ representation of the prebaryon constituents of $\phi$. Note that $d_\phi = 16$ for $y_{_{88}}$, and $d_\phi = 6$ for $y_{_{6\bar 6}}$, $y_{_{7/6}}$, $y_{_{\cal Q}}$.

Some of the di-prebaryons may develop VEVs, especially if their constituent prebaryons 
belong to higher SM gauge representations or have large Yukawa couplings in Eq.~(\ref{eq:Apsipsi}). 
VEV formation and the size of the VEVs is also sensitive to  
other interactions that may be present near the $\Lambda_{\rm pre}$ scale, such as heavy gauge bosons 
associated with extensions of the SM gauge group.
If a di-prebaryon acquires a VEV, then its two prebaryon constituents form a Dirac fermion that acquire 
a vectorlike mass. When the composite Dirac fermion carries color we refer to it as a Vquark, and otherwise as Vlepton.
We label the Vquarks whose components appear in (\ref{eq:largeYuk}) as $Q_{88}$, $Q_{6\bar 6}$,  $Q_{7/6}$ and  $\cal Q$. Their masses can be read off (\ref{eq:largeYuk}):  for example, 
the Vquark  ${\cal Q}$ gets a  mass  $m_{\cal Q} = y_{_{\cal Q}}   \langle \phi_{_{\cal Q}} \rangle$.

%%%%%%%%%%%
 \begin{figure}[t!]
  \begin{center} \vspace*{-.2cm}
 \includegraphics[width=0.99\textwidth, angle=0]{ 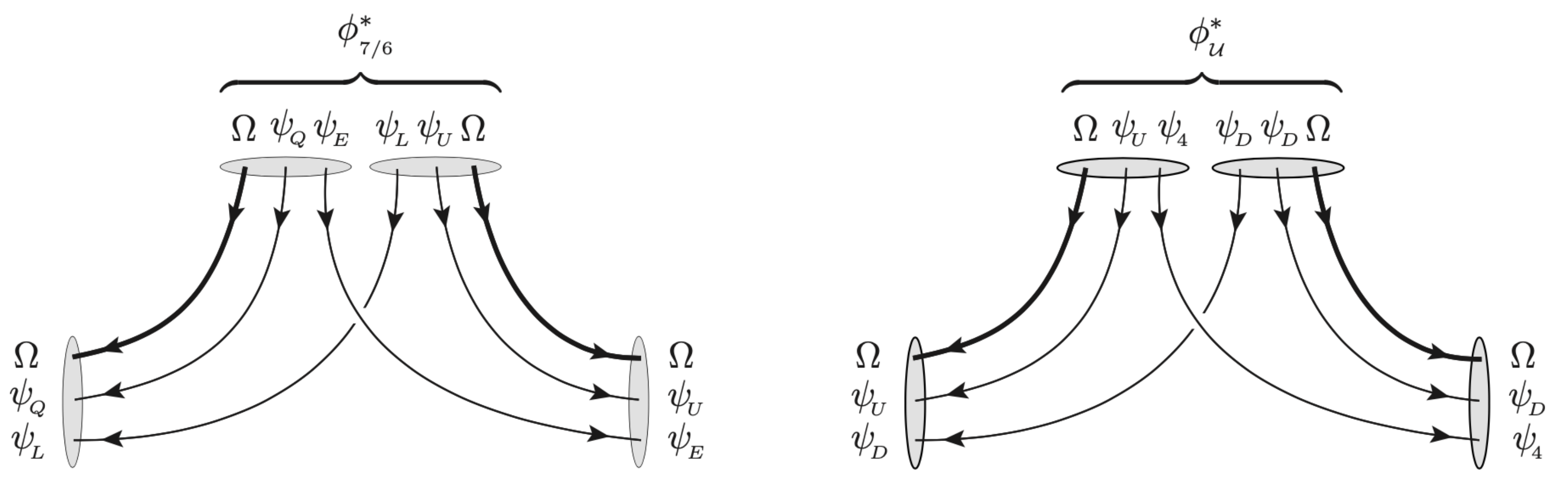}   
\vspace*{-0.1cm}
   \caption{Yukawa couplings   of the down-type weak-singlet Vquarks of components $\Omega_{QL }^{ (3, 1) }$, $ \Omega_{UD}^{ (3, 1)} $, $\Omega_{UE}$ and $\Omega_{D4}$ 
 to the composite scalars  $ \phi_{_{7/6}}$  (left diagram) and $\phi_{_{\cal U}}$  (right diagram).    }
  \label{fig:nonplanar}
  \end{center}
\end{figure}
%%%%%%%%%%%%

The second-order phase transition to nonzero VEVs for the composite scalars \cite{Bardeen:1993pj}  implies that the size of the VEVs depends on the coupling strength in the attractive channels.
Consequently, a quantitative estimate of the Vquark mass spectrum is 
hampered by the lack of information regarding how close to the critical value is the binding due to premeson exchange,
and also regarding other interactions near the compositeness scale. 
Note though, that the VEV of  $\phi_{\! _{88}}$ is larger than that of $\phi_{\! _{6\bar 6}}$ due to the higher color representation of the prebaryon constituents. This has an effect on the vectorlike mass that is at least partially compensated by the larger Yukawa coupling of $\phi_{\! _{6\bar 6}}$, since  Eq.~(\ref{eq:y0})  implies  $y_{\! _{6\bar 6}} = \sqrt{8/3} \, y_{_{88}}$. We assume in what follows that the Vquarks of higher color representations have roughly the same  mass, $m_{_{88}} \approx m_{_{6\bar 6}}$.

Vquarks of SM charges  $(3,3, -1/3)$, $(3,1, -4/3)$ and $(3,1, +2/3)$ 
obtain Dirac masses from the following Yukawa couplings between pairs of chiral prebaryons and 
the corresponding di-prebaryons:
\be
- y_{_{33}}  \phi_{_{33}}^*   \Omega_{QQ }^{ (\overline 3, 3) }   \Omega_{QL }^{ (3, 3) } 
- y_{_{4/3}}   \phi_{\! _{4/3}}^*   \Omega_{DE}  \Omega_{UU}
- y_{_{\cal U}}   \phi_{_{\cal U}}^*   \, \Omega_{DD}   \Omega_{U4} ~.
\label{eq:largeYukInterm}
\ee
These Vquarks are labelled $Q_{33}$ for the one which is a color-triplet and a weak-triplet, $Q_{4/3}$ for the 
one of hypercharge $-4/3$, and ${\cal U}$ for the one of the same charges as the SM weak-singlet up-type 
quarks.
We assume that the five di-prebaryons discussed so far that are bound states of color-triplet prebaryons have VEVs,
and for simplicity we take them to be roughly degenerate in mass,
$m_{_{\cal Q}} \approx m_{_{7/6}} \approx m_{_{33}} \approx m_{_{4/3}}  \approx  m_{_{\cal U}}$,
and lighter by a factor of order 3 than $Q_{88}$.

There are two remaining Vquarks, both of charges $(3, 1,-1/3)$. The associated di-prebaryons have the 
weakest binding induced by electroweak gauge boson exchange, so
it is consistent to assume that they do not acquire VEVs. Nevertheless, the down-type weak-singlet Vquarks 
acquire masses from preon-swap diagrams such as those shown in Figure~\ref{fig:nonplanar}. The mass matrix for the 
corresponding prebaryons arises from the following Yukawa couplings:
\be
- \left( \Omega_{QL}^{ (3, 1) }  \, , \, \Omega_{UD}^{ (3, 1) }   \right)    \left( \ba{cc}   
y_{_{7/6}}^\prime  \, \phi_{_{7/6}}^*  \; \;   &   \; \;  y_{_{\cal Q}}^\prime  \, \phi_{_{\! \cal Q}}^*  \\     y_{_{4/3}}^\prime  \, \phi_{_{4/3}}^*  \; \;  &  \; \;  y_{_{\cal U}}^\prime  \, \phi_{_{\cal U}}^*      \ea  \right)   \left( \ba{c}  \Omega_{UE}  \\  \Omega_{D4}  \ea  \right)    ~~.
\label{eq:D1D2Matrix}
\ee
Since the diagrams of Figure~\ref{fig:nonplanar} are nonplanar, the primed Yukawa coupling constants 
displayed here are suppressed by $1/N$ ($N =15$) compared to the corresponding 
couplings from (\ref{eq:largeYuk}) or (\ref{eq:largeYukInterm}), {\it e.g.}, $y_{_{7/6}}^\prime/y_{_{7/6}}  = O(1/15)$. 
After replacing the four above di-prebaryons by their VEVs, and diagonalizing the mass matrix, 
two physical Vquarks, labelled ${\cal D}_1$ and ${\cal D}_2$, are obtained. We do not expect the phases of the primed Yukawa couplings to be aligned, and thus there would not be a large hierarchy between the ${\cal D}_1$ and ${\cal D}_2$ masses even if the four VEVs were nearly degenerate.

%%%%%%%%%%%
 \begin{figure}[t!]
  \begin{center} \vspace*{-.2cm}
 \includegraphics[width=0.55\textwidth, angle=0]{ 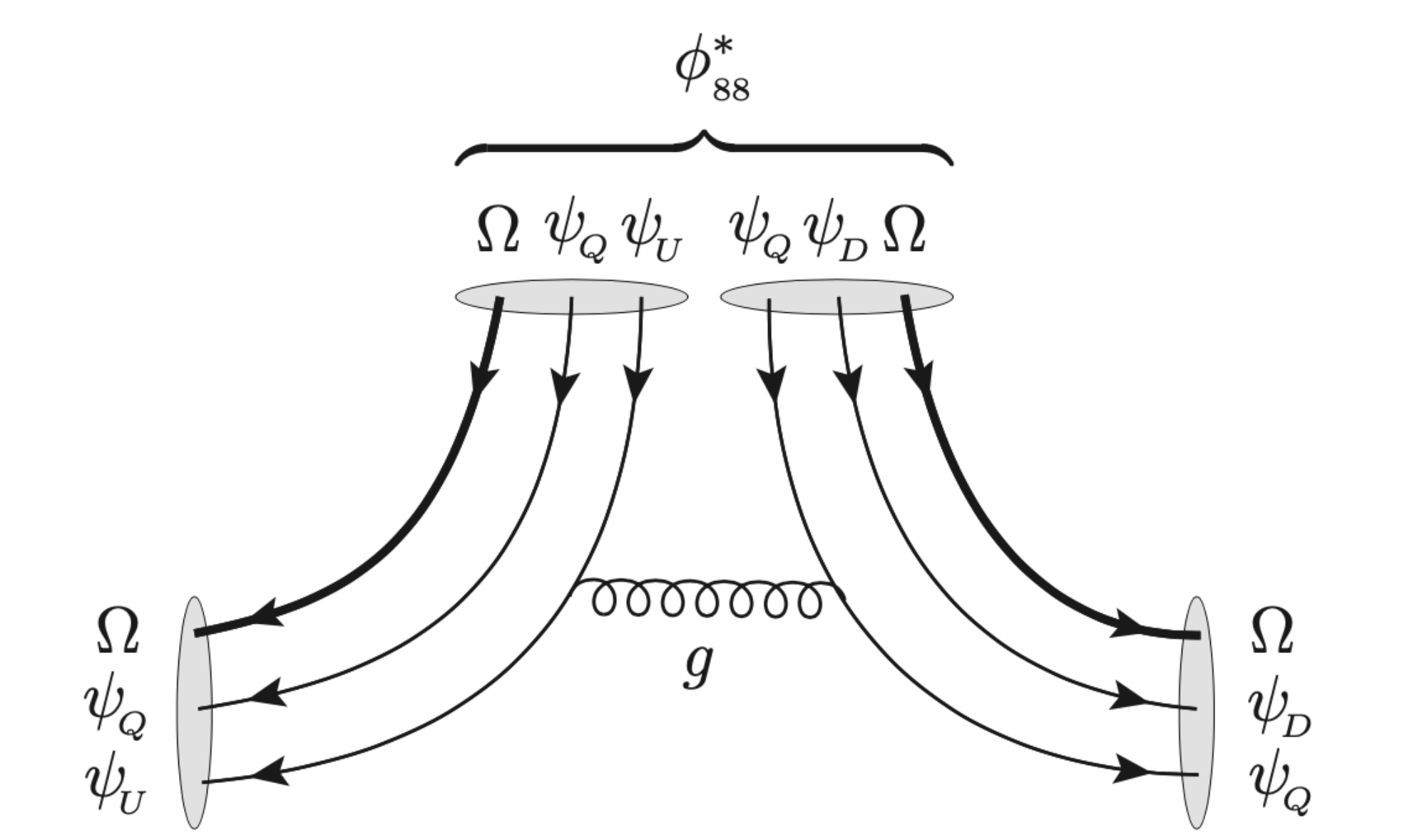}   
\vspace*{-0.2cm}
   \caption{Yukawa coupling of the weak-doublet Vlepton of components $  \Omega_{QU }^{ (1, 2) }  $ and   $  \Omega_{QD }^{ (1, 2) }  $
   to the composite scalar  $ \phi_{_{88}} $.      }
  \label{fig:QUQD}
  \end{center}
\end{figure}
%%%%%%%%%%%%

The di-prebaryons formed of color-singlet prebaryons are more weakly bound than those
formed of color-triplet prebaryons, and thus do not acquire VEVs. The Vleptons obtain masses through a few mechanisms, as briefly described in  \cite{Dobrescu:2021fny}. 
Let us first discuss the prebaryons of SM charges  $( 1 , 2, -1/2)  $, $\Omega_{L4}  $ and $\Omega_{QU}^{ (1, 2) }$,
which  together with the ones of charges $( 1 , 2, +1/2)  $,  $\Omega_{LE}$ and $  \Omega_{QD }^{ (1, 2) }   $, 
have a $2 \times 2$ mass matrix  due to the following Yukawa couplings:
\be
- \left( \Omega_{L4} \, , \, \Omega_{QU}^{ (1, 2) }   \right)    \left( \ba{cc}   
0  &   \; \;  y_{_{\cal Q}}^\prime  \, \phi_{_{\! \cal Q}}^*  \\   y_{_{7/6}}^\prime  \, \phi_{_{7/6}}^*  \; \;  &  \; \;  y^\prime_{_{88}}  \,  \phi_{_{88}}^*  + y^\prime_{_{6\bar 6}} \,   \phi_{_{6\bar 6}}^*     \ea  \right)   \left( \ba{c}  \Omega_{LE} \\    \Omega_{QD}^{ (1, 2) }      \ea  \right)    ~~.
\label{eq:VleptonMatrix}
\ee
The coupling to $ \phi_{_{88}}$ introduced here requires a gluon exchange (see Figure~\ref{fig:QUQD}), 
so it is smaller than the coupling  $y_{_{88}}$ of $\phi_{_{88}}$ to its constituents, the color octets  $\Omega_{QU }^{ (8, 2) }  $ and   $  \Omega_{QD }^{ (8, 2) }  $.  As a rough estimate, 
\be
y^\prime_{_{88}}  \sim  y_{_{88}} \,
\frac{\alpha_s}{4\pi}  \, \ln \left( \frac{ \Lambda_{\rm pre}^2 }{ m_{_{88}}^2  } \right)   ~~.
\label{eq:yplog}
\ee
The coupling to $ \phi_{_{6\bar 6}}$ displayed in (\ref{eq:VleptonMatrix}) arises from a 
nonplanar diagram that interchanges the $\psi_Q$ and $\psi_U$ preons
(similar to the diagrams of Figure~\ref{fig:nonplanar}), so we expect  $y^\prime_{_{6\bar 6}} /y_{_{6\bar 6}}  \sim O(1/N)$.  
Similarly, the couplings to $ \phi_{_{7/ 6}}$ and $ \phi_{_{\cal Q}}$   arise from nonplanar diagrams, 
so that $y^\prime_{_{7/ 6}} /y_{_{7/ 6}}  \sim y^\prime_{_{\cal Q}} /y_{_{\cal Q}}  \sim  O(1/15)$.
Replacing the scalars by their VEVs in (\ref{eq:VleptonMatrix}) gives a mass matrix with eigenstates 
representing two weak-doublet Vleptons, ${\cal L}_1$ and ${\cal L}_2$. 
Since $y^\prime_{_{88}}  \gg y^\prime_{_{\cal Q}}, y^\prime_{_{7/6}}$, the heaviest Vlepton (${\cal L}_2$) has a mass 
\be
m_{{\cal L}_2}  \approx  \left| y^\prime_{_{88}}  \langle \phi_{_{88}} \rangle^*  + y^\prime_{_{6\bar 6}} \,   \langle \phi_{_{6\bar 6}} \rangle^* \right|  ~~.
\label{eq:mL2}
\ee
The lighter weak-doublet Vlepton (${\cal L}_1$) has a mass given by a seesaw formula,
\be
m_{{\cal L}_1} = y^\prime_{_{\cal Q}} \, y^\prime_{_{7/6}} \,  \frac{ \langle \phi_{_{\cal Q}} \rangle  \langle \phi_{_{7/6}} \rangle }{ m_{{\cal L}_2} }    ~~.
\ee
Up to small corrections due to the mass mixing, the 4-component fermion ${\cal L}_2$ is formed of the prebaryons 
$\Omega_{QU}^{ (1, 2) }$   and   $\Omega_{QD}^{ (1, 2) }$. As these are left-handed Weyl fermions of opposite charges, 
we choose the sign of the hypercharge of   ${\cal L}_2$ (and also of ${\cal L}_1$) to be $-$, implying  that the left-handed component of ${\cal L}_2$ is
$\Omega_{QU}^{ (1, 2) }$ while the right-handed component is $\overline \Omega_{QD}^{ (1, 2) }$, 
both carrying SM charges $( 1 , 2, -1/2)  $.
In the same approximation 
where mixing is neglected, the left- and right-handed components of  ${\cal L}_1$ are 
$\Omega_{L4}$ and  $\overline \Omega_{LE}$, respectively.

%%%%%%%%%%%
 \begin{figure}[t!]
  \begin{center}
   \vspace*{-4.2cm}  
    \hspace*{-3.5cm}  
 \includegraphics[width=0.7\textwidth, angle=0]{ 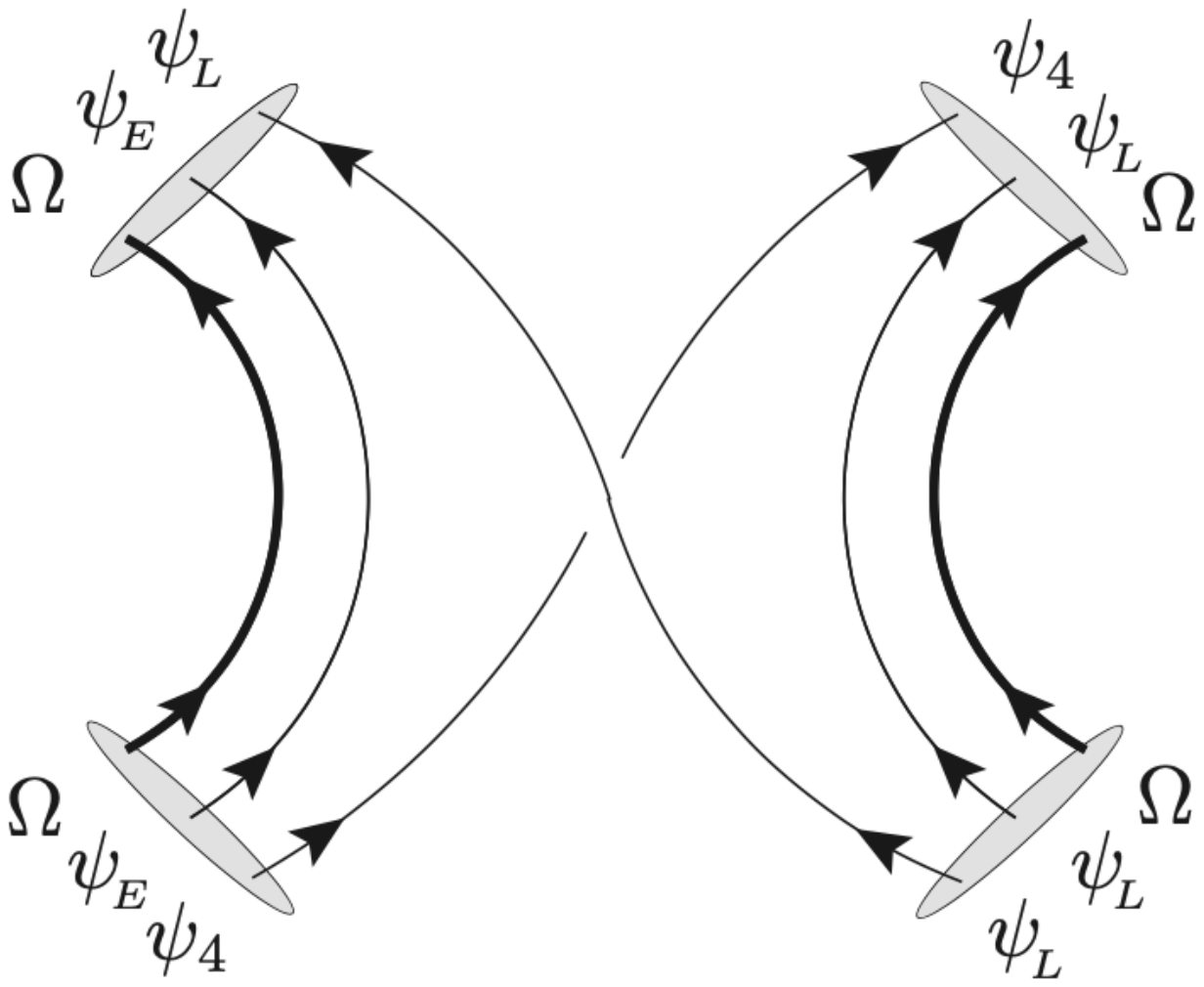}   
 \hspace*{-2.5cm}  
  \parbox[b]{4cm} {  \includegraphics[width=0.53\textwidth, angle=0]{ 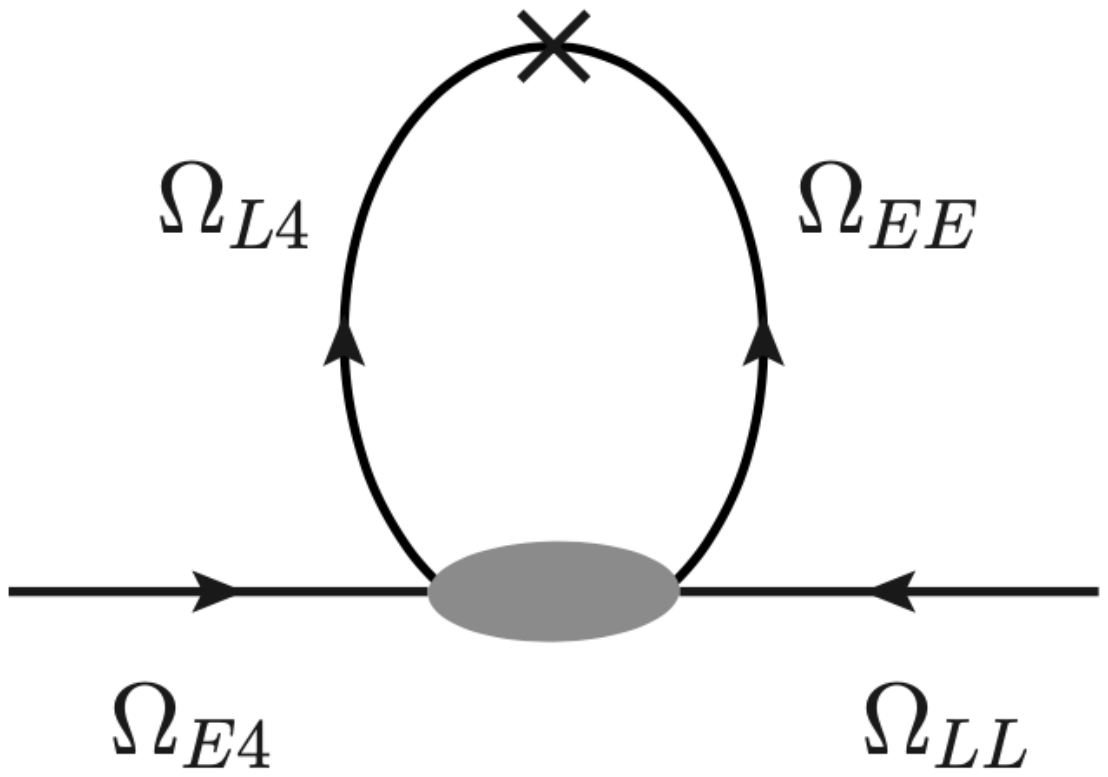}  \\ [1cm] }
 \vspace*{-4.5cm}
\caption{{\it Left:}  Four-fermion operator 
$  \left( \Omega_{ E4} \, \Omega_{LL}   \right)  \left( \overline \Omega_{ L4}  \, \overline \Omega_{LE} \right) $
generated by $SU(15)_{\rm p}$ dynamics at the scale  $\Lambda_{\rm pre}$, through a nonplanar diagram. 
{\it Right:} One-loop mass for the weak-singlet Vlepton  ${\cal E}$;
the~\scalebox{3}[1.5]{\color{gray}  $\bullet$}~represents   the 4-fermion operator from the left diagram, and the $\times$ represents the 
    mass of the weak-doublet Vlepton  ${\cal L}_1$.    }
  \label{fig:4fermion}
  \end{center}
\end{figure}
%%%%%%%%%%%%

There is also a Vlepton of charges $(1 , 1, -1) $, labelled ${\cal E}$, with  left- and right-handed 
components given by $ \Omega_{ E4}  $  and  $  \overline \Omega_{LL}  $. 
At the scale  $\Lambda_{\rm pre}$,  $SU(15)_{\rm p}$ dynamics generates a 4-prebaryon operator  
$  \left( \Omega_{ E4} \,  \Omega_{LL}   \right)  \left( \overline \Omega_{ L4} \,  \overline \Omega_{LE} \right) $,
through the nonplanar diagram shown on the left side of Figure~\ref{fig:4fermion}. An order of magnitude estimate 
gives the coefficient of the operator to be $2 (4\pi)^2/(N \Lambda_{\rm pre}^2)$, where the prefactor of 2 arises from the 
interchange of two $\psi_L$ preons in $\Omega_{LL}  $. 
A mass for the ${\cal E}$ Vlepton then is induced by the loop diagram shown in Figure~\ref{fig:4fermion}, and is parametrically given by
\be
m_{\cal E} \approx  4 \,  \frac{m_{ {\cal L}_1} }{N}    ~~,
\label{eq:Eloop}
\ee
where an extra factor of 2 arises from the doublets running in the loop.
   
The 12 charged vectorlike fermions and the parametric expressions of their masses are listed in Table \ref{table:Vquarks}.
The heaviest fermion, $ \VQ_{88}$, of components  $\Omega_{QU }^{(8,2)}  $  and $ \Omega_{QD}^{(8,2)} $,
appears on the first row, and the other Dirac fermions are listed in decreasing mass order, with some adjacent rows 
displaying fermions (such as $ {\cal Q}$   and $ \VQ_{4/3} $) that are roughly degenerate in mass.
Recall that the only other composite fermions lighter than $\Lambda_{\rm pre}$ are three generations of chiral quarks and leptons, as well as six gauge-singlet Weyl fermions, $\Omega_{ij}$ with $1 \le i < j \le 4$
(note that $\Omega_{43} = - \Omega_{34}$).  

%%%%%%%%%%%%%%%%%%%%%
\begin{table}[tb]
\begin{center}
\renewcommand{\arraystretch}{1.4}
\begin{tabular}{|c|c|c|c|}\hline    
 \parbox[t]{1.7cm}{vectorlike\\ [-0.1cm]   \hspace*{0.1cm} fermion \\ [-0.4cm]  }
& \parbox[t]{2.25cm}{ \ component \\ [-0.1cm]  \hspace*{0.1cm}  LH  \  , \hspace*{-0.05cm} RH  \\ [-0.3cm]  }  & 
 \parbox[t]{4.4cm}{  \hspace*{0.1cm}  \\ [-0.3cm]  $SU(3)_c  \! \times  \! SU(2)_W  \!  \times  \!  U(1)_Y$    }   &    \parbox[t]{0.9cm}{  \hspace*{0.1cm}  \\ [-0.35cm]   mass  }
\\ \hline \hline
$ \VQ_{88}  $ &  $  \Omega_{QU }^{(8,2)}  $  ,  $ \overline \Omega_{QD}^{(8,2)}  $     &   $( 8 , 2, -1/2) $    &     $y_{_{88}} \,  \langle \phi_{_{88}} \rangle $   
 \\ \hline
$ \VQ_{6 \bar 6}  $ &  $\Omega_{QQ }^{(6, 1)}  $ , $  \overline \Omega_{UD }^{(\overline 6, 1)} $     &     $( 6 , 1, +1/3) $          &     $y_{_{6 \bar 6}}  \,  \langle \phi_{_{6 \bar 6}} \rangle $
\\ \hline
$ \VQ_{33}  $ &     $\Omega_{QL }^{(3,3)} $   ,  $  \overline \Omega_{QQ }^{(\overline 3, 3)}   $             &  $ ( 3 , 3, -1/3)   $      &    $y_{_{33}}  \,  \langle \phi_{_{33}} \rangle $
\\ \hline
$ \VQ_{7/6}   $ &         $ \Omega_{Q E }  $   ,   $  \overline \Omega_{U L } $    &  $(  3 , 2, +7/6) $  &      $y_{_{7/6}}   \,  \langle \phi_{_{7/6}} \rangle $    
\\ \hline
$ {\cal Q}    $ &    $ \Omega_{Q4 }  $   ,   $  \overline \Omega_{D L } $       &     $( 3 , 2 , +1/6) $ &      $y_{_{\cal Q}}   \,  \langle \phi_{_{\cal Q}} \rangle $    
\\ \hline
$ \VQ_{4/3}   $ &    $  \Omega_{UU } $    ,    $  \overline \Omega_{DE } $        &     $( 3 , 1, -4/3) $ &       $y_{_{4/3}}  \,  \langle \phi_{_{4/3}} \rangle $    
 \\  \hline  
    $  {\cal U}   $ &   $ \Omega_{ DD}  $  ,  $  \overline \Omega_{U 4 }  $     &     $( 3 , 1, +2/3) $ &                    
      $y_{_{\cal U}}  \,  \langle \phi_{_{\cal U}} \rangle $    
 \\ \hline
   $  {\cal L}_2    $ &  $  \Omega_{QU }^{ (1,2) }  $ ,    $  \overline \Omega_{QD }^{ (1,2) }  $       &     $( 1 , 2, -1/2)  $   &     
   $ \left| y_{_{88}}^\prime  \,  \langle \phi_{_{88}} \rangle^*  + y_{_{6\bar 6} }^\prime \langle \phi_{_{6\bar 6} } \rangle^*  \right| $ 
\\ \hline 
   $  {\cal D}_2   $ &    $  \Omega_{ QL }^{(3, 1)} $   ,  $  \overline \Omega_{U E }  $  &       &       
   $\sim y_{_{7/6}}^\prime \langle \phi_{_{7/6}} \rangle $   
\\  [-4mm]
   & &  $( 3 , 1, -1/3) $  & 
\\  [-4mm]
     $  {\cal D}_1   $ &     $  \Omega_{ UD}^{(3,1)}  $   ,  $ \,   \overline  \Omega_{ D4 }  $     &     &        
     $\sim  y_{_{\cal U}}^\prime  \,  \langle \phi_{_{\cal U}} \rangle$   
\\ \hline
   $  {\cal L}_1    $ &  $    \Omega_{ L 4 }    $  ,  $   \overline \Omega_{LE }    $     &  $  ( 1, 2, -1/2)  $  &            $ \displaystyle   \left.  y_{_{\cal Q}}^\prime \,  y_{_{7/6}}^\prime    \langle \phi_{_{\cal Q}} \rangle   \frac{  
      \langle \phi_{_{7/6}} \rangle }{ m_{  {\cal L}_2 } }_{\rule{0mm}{7mm} }    \rule{0mm}{8mm} \right. $ 
\\ \hline
        $  {\cal E}   $ &  $ \Omega_{E4 }  $   ,  $  \overline \Omega_{LL }    $     &  $  ( 1, 1, +1)  $  &     $ \displaystyle   \left. 
       \frac{4}{N}_{  \left. \rule{0mm}{6mm}  \right. } m_{ {\cal L}_1}     \rule{0mm}{8mm} \right. $ 
 \\ \hline
 \end{tabular}
\caption{Composite vectorlike quarks and leptons, and their representations under the SM gauge group.
Left-handed components of the vectorlike fermions are prebaryons, while right-handed components
are conjugates of other prebaryons. For example, the vectorlike quark ${\cal U}$ has a left-handed component
given by the $\psi_{\! _D} \psi_{\! _D} \Omega \equiv \Omega_{DD}$ prebaryon,
while its right-handed component is the conjugate of the $\psi_{\! _U} \psi_{\! _4}  \Omega \equiv \Omega_{U 4 }$ prebaryon.
The table rows are ordered according to the fermion masses, with the heaviest one at the top, but some of the Vquarks are roughly degenerate.
Besides vectorlike fermions, shown here, there are 3 SM generations of composite chiral quarks and leptons. \\ [-1cm]
}
\label{table:Vquarks}
\end{center}
\end{table}

The current lower limits on Vleptons and Vquarks, set by the CMS and ATLAS experiments assuming decays 
into a SM boson and a third generation fermion, are given by $m_{\cal E}  > 150$ GeV, $m_{{\cal L}_1}  > 1.0$ TeV \cite{CMS:2022nty},  $m_{\cal U}, m_{{\cal D}_1}   > 1.5$ TeV \cite{ATLAS:2018ziw}.
As an example of a mass spectrum for the vectorlike fermions, we take the mass of the lightest weak-doublet Vlepton to be $m_{{\cal L}_1} = 1$ TeV,
and the compositeness scale $\Lambda_{\rm pre} = O(5 \times 10^4)$ TeV (this value will be justified in Section \ref{sec:perturbativity}), and use the mass relations discussed above. The weak-singlet Vlepton has then a mass $m_{\cal E} = 260$ GeV, and the heaviest weak-doublet Vlepton mass is roughly $m_{{\cal L}_2} \approx 4$ TeV, where we assumed that the first term in Eq.~(\ref{eq:mL2}) is the dominant one.
The down-type Vquark masses are $m_{{\cal D}_{1,2}} \approx O(2)$ TeV, the five intermediate-mass Vquarks 
have masses of order  $m_{_{\cal Q}}  \approx 30$ TeV, while the color octet and sextet Vquarks have masses  
$m_{_{88}} \approx m_{_{6 \bar 6}} \approx  90$ TeV.
Note that the masses of these Vquarks depend logarithmically on $\Lambda_{\rm pre}$ through the Yukawa coupling in
Eq.~(\ref{eq:y0}), and the estimate is self-consistent as long as they are not larger than $\Lambda_{\rm pre}$. 

Besides the gauge-singlet di-prebaryons discussed above
($\phi_{_{88}}$, $\phi_{_{6\bar 6}}$, $\phi_{_{33}}$, $\phi_{_{7/6}}$, $\phi_{_{4/3}}$, $\phi_{_{1/6}}$, $\phi_{_{2/3}}$),
there is a scalar labelled $\phi_{\rm _M}$ that arises as an $\Omega_{34}  \Omega_{34} $
bound state \cite{Dobrescu:2021fny}. Its VEV may be comparable to $\langle \phi_{_{88}} \rangle$, and generates a large Majorana mass for the $\Omega_{34}$ prebaryon.
As the eight composite scalars listed above have VEVs that break the chiral symmetries of various prebaryons, 
their angular degrees of freedom behave like Nambu-Goldstone bosons. Each of these pseudoscalars acquires a mass
much smaller than the corresponding VEV, induced by higher-dimensional operators such as 
$\Omega_{QU }^{(8,2)}  \Omega_{QD}^{(8,2)}  \overline \Omega_{QQ }^{(6, 1)} \overline \Omega_{UD }^{(\overline 6, 1)}  $.
The radial degrees of freedom of the composite scalars have masses comparable to their VEVs,
roughly in the $10-100$ TeV range. 

In addition to the gauge-singlet composite scalars described above, there are weak-doublet 
di-prebaryons. Assuming $\lambda_{43} >  \lambda_{ij}$ for $i,j = 2,3$, the largest VEV of the 
composite weak doublets is that of the $H_u = \Omega_{Q3}  \Omega_{U4}$ scalar,
which is responsible for the up-type SM quark masses.
The $H_d  = \Omega_{Q3}  \Omega_{D4}$ scalar acquires a VEV that is about an order of magnitude smaller than that of 
$H_u$. Other weak-doublet scalars also form, but it is consistent to assume that their VEVs vanish. Among these, the 
$H_{i j} = \Omega_{Q i}  \Omega_{U j}$ di-prebaryons for $i,j = 1,2,3$ are likely to be substantially lighter than 
$\Lambda_{\rm pre}$.

\medskip

%%%%%%%%%%%%%%%%%%%%%%%%%%%%%%%%%%%%%%%%%%%%%%%%%%%%%%%%
%%%%%%%%%%%%%%%%%%%%%%%%%%%%%%%%%%%%%%%%%%%%%%%
\section{Constraints from proton decay searches }
\label{sec:decay}\setcounter{equation}{0}

\begin{figure}[t!]
  \begin{center}
  \vspace*{-1.1cm}  
%    \hspace*{-3.5cm}  
 \includegraphics[width=0.6\textwidth, angle=0]{ 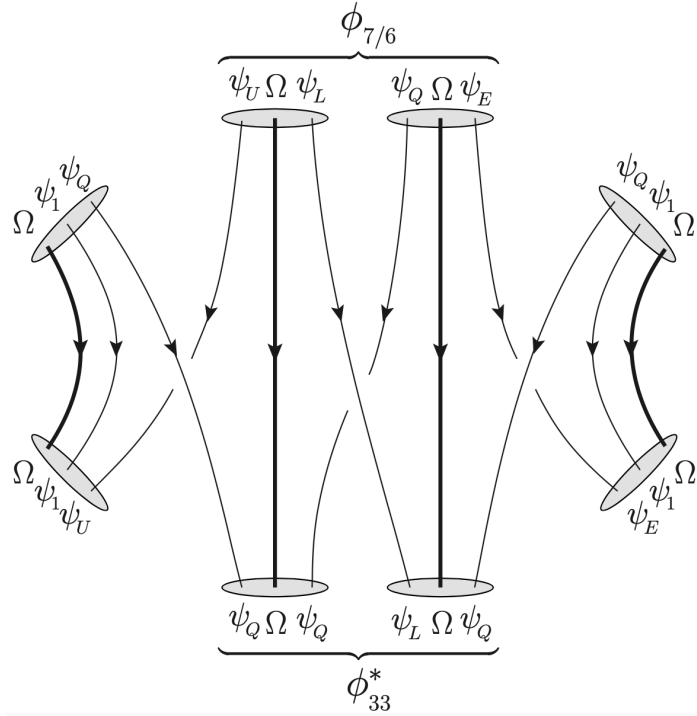}     
 \vspace*{-1.3cm}
\caption{Dimension-8 baryon-number violating operator (\ref{eq:dim8})
generated by $SU(15)_{\rm p}$ dynamics at the scale  $\Lambda_{\rm pre}$, through a nonplanar diagram. 
Each line corresponds to a certain preon exchanged between two prebaryons, which are represented by the gray ovals.
The scalars $\phi_{ _{7/6}}$ and $\phi_{_{33}}$ are di-prebaryons that acquire VEVs.
   }
  \label{fig:QQQLnoGUT}
  \end{center}
\end{figure}

Let us now discuss proton decay operators generated at the compositeness scale, $\Lambda_{\rm pre}$.  Consider the gauge-invariant 8-prebaryon operator
\be
\frac{\widetilde C_8}{2 \Lambda_{\rm pre}^{8}}
\left( \Omega_{QE}  \,  \Omega_{UL}  \right) 
\left(  \overline  \Omega_{QQ}^{(\overline  3,3)}  \,  \overline  \Omega_{QL}^{(3,3)}   \right)   
\left(  \Omega_{Q1} \Omega_{Q1}   \right)  
\left( \overline  \Omega_{U1}  \overline  \Omega_{E1}  \right)       ~~.
\label{eq:QQQLnoGUT}        
\ee
This operator is generated by the $SU(15)_{\rm p}$ dynamics, as shown in Figure \ref{fig:QQQLnoGUT}.
We impose that  the $SU(3)_c\times SU(2)_W$ indices of the $\psi$ preons are contracted such that 
the $ \Omega_{QE}   \Omega_{UL}$ bilinear is a gauge singlet. 
Given that  the di-prebaryon $\phi_{ _{7/6}}$ is an $\Omega_{QE} \Omega_{UL} $ bound state,
the $ \Omega_{QE}  \Omega_{UL}$ bilinear can be replaced by $\tilde c_{ _{7/6}} \Lambda_{\rm pre}^{2} \phi_{ _{7/6}}$.
The coefficient  $\tilde c_{ _{7/6}}$   can be written as 
$c_\phi \, y_{_{7/6}}  /(4 \pi)^2$ with $c_\phi$ presumably of order unity, 
because there is a loop suppression 
when the  $\Omega_{QE}$ and $\Omega_{UL} $  legs of the operator (\ref{eq:QQQLnoGUT}) merge into 
a scalar through the $y_{_{7/6}} $ Yukawa coupling from (\ref{eq:largeYuk}).  Analogously,  
we impose that the $ \overline  \Omega_{QQ}^{(\overline  3,3)}  \,  \overline  \Omega_{QL}^{(3,3)}  $  bilinear is a SM gauge singlet, 
so  that it can be replaced 
by $c_\phi \, y_{_{33}}^*  (4 \pi)^{-2} \, \Lambda_{\rm pre}^{2} \phi_{_{33}}^\dagger$. Thus, operator (\ref{eq:QQQLnoGUT}) gives rise to a dimension-8 operator,
\be
\frac{\widetilde C_8 \, c_\phi^2 \, y_{_{7/6}} y_{_{33}}^* }{ 2 (4 \pi)^4  \Lambda_{\rm pre}^{4}} \,
\phi_{ _{7/6}} \phi_{33}^\dagger
\left(  \Omega_{Q1} \Omega_{Q1}   \right)  
\left( \overline  \Omega_{U1}  \overline  \Omega_{E1}  \right)       
~~.
\label{eq:dim8}
\ee

%%%%%%%%%%%
%%%%%%%%%%%%

The diagram displayed in Figure \ref{fig:QQQLnoGUT} is nonplanar, such that $C_8$ is suppressed by powers of $1/N$ and also of $1/N_c$, where $N = 15$, and $N_c = 3$ is the number of QCD colors.
To count powers of $1/N$, note that there are 8 in-going external lines representing $\psi$ preons, but only five of them have arbitrary
$SU(15)_{\rm p}$ indices. Thus, the dimensionless coefficient $\widetilde C_8$ is suppressed by $1/N^3$.
Same result can be obtained by counting the number of nonplanar line pairs.
The powers of $1/N_c$ can be counted by noticing that there are two constraints on the $SU(3)_{c}$ indices of the incoming
preons, arising from the requirements that  $\Omega_{QE}\Omega_{UL} $ and 
$ \left(  \Omega_{Q1} \Omega_{Q1}   \right)   \overline  \Omega_{U1}  $ are color singlets. Thus, we can write 
\be
\widetilde C_8 = \frac{ C_8}{N^3  N_c^2}   ~~.
\label{eq:Nfactors}
\ee
The dimensionless coefficient $C_8$ does not depend on $N$ or $N_c$, but we cannot 
assess whether it is of order one given the lack of understanding of the chiral $SU(15)_{\rm p}$  dynamics.

Identifying the chiral prebaryons by the corresponding SM fermions, and replacing the scalars by their VEVs in (\ref{eq:dim8}),
we obtain the following dimension-6 baryon-number violating operator:
\be
C_8 \, c_\phi^2  \;  \frac{ m_{_{7/6}} m_{_{33}} }{ 2 (4 \pi)^4 N^3 N_c^2 \, \Lambda_{\rm pre}^{4}} \,
\left(  q_L^1  q_L^1   \right)   \left(  \overline u^c \,  \overline e^c  \right) 
 ~~,
 \label{eq:qqql3}
\ee
where $q_L^1$ is the first generation quark doublet,  $m_{_{7/6}} = y_{_{7/6}}  \,  \langle \phi_{ _{7/6}}   \rangle  $ is the mass of the $ \VQ_{3,2}   $ Vquark (see Table \ref{table:Vquarks}), and 
$m_{_{33}}  = y_{_{33}} \langle  \phi_{_{33}}  \rangle$  is the mass of the $ \VQ_{3,3}   $ Vquark. 
Same baryon-number violating operator gets contributions from 
diagrams analogous to the one in Figure \ref{fig:QQQLnoGUT} but with 
$\phi_{ _{7/6}}  \phi_{ _{33}}^*$ replaced by $\phi_{_{4/3}}  \, \phi_{_{88}}^*$ or $\phi_{_{4/3}} \,  \phi_{_{6\overline 6}}^*$.
These, however, are of order $1/(N N_c)^3$ and we will ignore them.
Contracting the $SU(2)_W$ indices in the operator (\ref{eq:qqql3}) gives
\be
\frac{1}{M_{qque}^2}   \left(  u_L  d_L  \right)   \left(  u_R  \, e_R  \right)   ~~,
\label{eq:uLdLuReR}
\ee
where  the mass scale of this dimension-6 operator is 
\be
M_{qque} = \frac{ 16 \pi^2  N^{3/2} \, N_c \, \Lambda_{\rm pre}^2}{  c_\phi \,  ( C_8 \,  m_{_{7/6}} m_{_{33}} )^{1/2}}    ~~.
\ee
The dominant proton decay mode induced by operator (\ref{eq:uLdLuReR}) is $p \to e^+ \pi^0$.
The hadronic matrix element for the proton decay induced by this operator 
can be written as 
\be
\langle \pi^0 |   \left(  u_L  d_L  \right) u_R \, | p \rangle_{q^2 = m_e^2} = \frac{1}{\sqrt{2}} P_R \left( W_0 (m_e)  
- W_1  (m_e) \,  q_\mu \gamma^\mu 
 \rule{0mm}{3.9mm} \right)  \! {\textsl u}_p  ~~,
 \label{eq:W0}
\ee
where $ {\textsl u}_p $ is the proton spinor operator, and $q_\mu$ is the electron 4-momentum. 
Form factors $W_0 (m_e)$ and $W_1 (m_e)$ 
have been computed on the lattice \cite{Yoo:2021gql} for $q^2 = 0$, which is a good approximation 
given how small $m_e/m_p$ is. The contribution proportional to   $W_1$ is also suppressed by $m_e/m_p$,
and can be neglected. The result for the first form factor is $W_0 (0) \approx - 0.16\,\,  {\rm GeV}^2  $ at a scale of 2 GeV.
If there are no other comparable contributions to the $p \to e^+ \pi^0$ process, the decay width of the proton is 
\be
\Gamma \! \left( p  \to  \pi^0  e^+  \right)  = \frac{m_p \, W_0^2 (0)}{64\pi \, |M_{qque}|^4}  
\left( 1 -  \frac{m_{\pi^0}^2 }{m_p^2}  \right)^{\! 2}   
\label{eq:pwidth}
\ee
which results in a proton lifetime 
\be
\tau \!\left( p  \to  \pi^0  e^+  \right)  = \frac{ 1.0 \times 10^{34} \,  {\rm yr}  }{ |C_8|^2 \,  |c_\phi|^4   }   \left( \frac{ \Lambda_{\rm pre}} {10^4 \;  {\rm TeV}} \right)^{\! 8}
 \left( \frac{1 \;  {\rm TeV}^2 }{  m_{ _{7/6}}  m_{_{33}}  }\right)^{\! 2}      ~~.
\ee
Assuming $|c_\phi| \, = \! 1$, 
the current experimental limit on the proton lifetime set by the Super-Kamiokande collaboration \cite{Super-Kamiokande:2020wjk}, 
$\tau \!\left( p  \to  \pi^0  e^+  \right)  > 1.6 \times 10^{34} \,  {\rm yr} $, implies
\be
\frac{ \Lambda_{\rm pre} }{  |C_8|^{1/4} }  > 1. 0 \times 10^4 \;  {\rm TeV}  
 \left(  \frac{ m_{ _{7/6}}  m_{_{33}}    }{1 \;  {\rm TeV}^{2}   }  \right)^{\! 1/4}   ~~.
\label{eq:const1}
\ee
As discussed in Section \ref{sec:below}, the lower limit on the Vquark masses $m_{ _{7/6}}$ and $m_{_{33}}$ is about 1.5 TeV if the 
decays include a SM boson and a quark. In that case $ \Lambda_{\rm pre} / |C_8|^{1/4}  > 1.3 \times 10^4 \;  {\rm TeV} $.
However, the lower limit on the Vquark masses is substantially relaxed if they have certain exotic decays, such as into three SM quarks \cite{Dobrescu:2016pda}, with large branching fractions. Nevertheless, there are other constraints on  $m_{ _{7/6}}$ and $m_{_{33}}$  
from the fact that the di-prebaryon VEVs are also responsible for the weak-doublet Vlepton masses (see Table \ref{table:Vquarks}), 
as we discuss in Section \ref{sec:perturbativity}.
Note that we have not included running effects \cite{Abbott:1980zj} on the coefficient of the proton-decay operator, but the size of those effects is much smaller here than in grand unified theories where the operator is generated at a scale that is higher by nine orders of magnitude.

There are also  two  8-prebaryon operators, similar to (\ref{eq:QQQLnoGUT}),
which yield contributions  proportional to $m_{_{4/3}} m_{_{88}}$ and  $m_{_{4/3}} m_{_{6\bar 6}}$ 
 to the baryon-violating operator (\ref{eq:uLdLuReR}). Those, however, are 
suppressed by and extra factor of $1/N_c$, and thus may be important only if the corresponding Vquarks are heavier.
Note that the interference between these contributions and (\ref{eq:uLdLuReR}) depends on unknown coefficients, 
and thus we do not know if they reduce or increase the proton width. 

Two other 8-prebaryon operators generated by the $SU(15)_{\rm p}$ dynamics,
\be
\frac{1}{\Lambda_{\rm pre}^{8}}
\left[
\widetilde C_8^\prime  \left(  \overline  \Omega_{QE}  \,   \overline \Omega_{UL}  \right) 
\left(    \Omega_{UU}  \,    \Omega_{DE}  \right)   
+
\widetilde C_8^{\prime\prime}  \left(  \overline  \Omega_{Q4}  \,   \overline \Omega_{DL}  \right) 
\left(    \Omega_{U4}  \,    \Omega_{DD}  \right)   
  \rule{0mm}{4mm}\right]   
\left( \overline  \Omega_{U1}  \overline  \Omega_{D1}  \right)     
\left(  \Omega_{Q1} \Omega_{L1}   \right)      ~~,
\hspace{1cm}
\label{eq:udQL}            
\ee
which have coefficients suppressed by $1/(N^3 N_c^2)$, lead to a different baryon-number violating operator:
\be
\frac{1}{M_{udql}^2}   \left(  u_R  d_R  \right) \left(  q_L^1  l_L^1  \right)   ~~.  
\label{eq:uRdRuLeL}
\ee
Writing $ C_8^\prime = \widetilde C_8^\prime N^3  N_c^2$ and analogously for 
$ C_8^{\prime \prime}$, the mass scale of the above dimension-6 operator is given by 
\be
M_{udql} = \frac{ 16 \pi^2  N^{3/2} \, N_c \, \Lambda_{\rm pre}^2}{  c_\phi \,  ( C_8^\prime \,  m_{_{4/3}}  m_{_{7/6}} + C_8^{\prime \prime} \,  m_{_{\cal U}}  m_{_{\cal Q}}   )^{1/2}}    ~~.
\ee
Note that $ C_8^{\prime}$ and $ C_8^{\prime \prime}$ are complex parameters, and their relative phase is unknown.
There are also some contributions to the operator (\ref{eq:uRdRuLeL}) proportional to $m_{_{33}}  m_{_{88}}$
or $m_{_{33}}  m_{_{6\bar 6}}$, but they are suppressed by $1/(N N_c)^3$. $SU(15)_{\rm p} $ dynamics cannot produce other 8-prebaryon operators that lead to proton-decay into first-generation SM fermions.  

Operator (\ref{eq:uRdRuLeL}) gives additional contributions to the $p \to e^+ \pi^0$ process, so that the left-hand side of the (\ref{eq:const1}) constraint is modified, and is given by
\be
  \left(  |C_8 + O(1/N_c) |^2 + \left| C_8^\prime + C_8^{\prime \prime} + O(1/N_c) \right|^2 \right)^{-1/8}  \Lambda_{\rm pre} 
\ee
in the particular case where the Vquark masses are equal. 
Operator (\ref{eq:uRdRuLeL}) also contributes to  $p \to \pi^+ \nu$, but the experimental limit for this process is weaker,
so the (\ref{eq:const1}) constraint remains the most stringent one. 

An 8-prebaryon operator analogous to (\ref{eq:QQQLnoGUT}) 
with $\left(  \Omega_{Q1} \Omega_{Q1}   \right)  \left( \overline  \Omega_{U1}  \overline  \Omega_{E1}  \right)$
replaced by 
$\left(  \Omega_{Q1} \Omega_{Q2}   \right)  \left( \overline  \Omega_{U1}  \overline  \Omega_{E2}  \right)$        
is generated as in Figure \ref{fig:QQQLnoGUT} but with one of the $\psi_1$ preons  replaced by a $\psi_2$.
That operator induces the $p \to K^0 \mu^+$ decay. Likewise, operators analogous to (\ref{eq:udQL}) 
involving second-generation fermions induce the $p \to K^+ \nu$ decay.
The constraints arising from these processes are, however, weaker than  (\ref{eq:const1}) for a few reasons:
the matrix elements involving kaons have smaller form factors than the $W_0$ for the pion that enters Eq.~(\ref{eq:pwidth}), 
the phase space is more suppressed, and the experimental limits are slightly weaker.
Other operators involving SM fermions, of the type $q_L \, q_L \, q_L \, l_L$ or $u_R \, u_R \, d_R \, e_R$ \cite{Weinberg:1979sa, Abbott:1980zj}, 
are much more suppressed in this model of compositeness because they violate the chiral symmetry of $\Omega$.

Searches for baryon-number violating decays of the neutron also impose constraints on $\Lambda_{\rm pre}$,
but those constraints are slightly less severe than the ones from proton decay searches. For example, 
the $n \to e^+ \pi^-$ process is induced by operator (\ref{eq:uRdRuLeL}), 
and the limit $\tau \!\left( n  \to  e^+ \pi^- \right)  > 5.3 \times 10^{33} \,  {\rm yr} $ implies 
$\Lambda_{\rm pre}/|C_8|^{1/4}   > 1. 2 \times 10^4 \;  {\rm TeV}$ for $m_{ _{7/6}} =m_{_{33}} =1.5$ TeV.

%\medskip

%%%%%%%%%%%%%%%%%%%%%%%%%%%%%%%%%%%%%%%%%%%%%%%%%%%%%%%%
%%%%%%%%%%%%%%%%%%%%%%%%%%%%%%%%%%%%%%%%%%%%%%%
\section{Maintaining  perturbativity of QCD up to $\Lambda_{\rm pre}$}
\label{sec:perturbativity}\setcounter{equation}{0}

At scales higher than the Vquark masses, the QCD coupling constant, $\alpha_s$, loses asymptotic freedom due to the large number of composite Vquarks. The RGE for $\alpha_s$ is
\be
\frac{d \alpha_s  (\mu)}{d \ln \mu} = - 2  \alpha_s^2  (\mu)  \,  b_3 (\mu)   ~~,
\ee
where the coefficient of the 1-loop $\beta$ function at a scale $\mu$ above the top quark mass is 
\be
b_3 (\mu) = 7 - \frac{2}{3} n_3(\mu)  - \frac{10}{3} n_6(\mu)  - 4 n_8(\mu)    ~~.
\ee
Here $n_3(\mu)$, $n_6(\mu)$ and $n_8(\mu)$ are the number of Dirac fermions transforming in the 3, 6, and 8 representations of $SU(3)_c$, respectively. For $\mu$ above the heaviest Vquark mass, $n_3(\mu)= 11$,   $n_6(\mu)= 1$, $ n_8(\mu)= 2$.

The SM value of $\alpha_s$ at the $m_t = 172$ GeV scale is $\alpha_s (m_t) \approx 0.1074$ \cite{Kniehl:2016enc}. 
If all color-triplet Vquarks have a mass $m_3 = 2$ TeV, and the higher-representation Vquarks 
have a mass $m_{_{88}} = 3 m_3$, then running $\alpha_s$ with $b_3 = 7$ for $\mu$ between $m_t$ and $m_3$,
and with $b_3 = -1/3$ for $m_3 < \mu < m_{_{88}} $ gives $\alpha_s (m_{_{88}}  )= 0.0708$ at the scale of the heaviest Vquarks. 
At larger scales, $b_3 = -35/3$ and the solution to the RGE is
\be
\frac{1}{\alpha_s (\mu)} = \frac{1}{\alpha_s (m_{_{88} })}  + \frac{35}{6\pi} \ln \left( \frac{\mu}{  m_{_{88} } } \right)   ~~.
\ee
Thus, $\alpha_s$ increases at larger $\mu$, and it reaches a nonperturbative value at a scale $\mu_\infty \approx  100$ TeV.
As this is below the lower limit on $\Lambda_{\rm pre}$ derived based on the proton lifetime in the previous section, it appears necessary to raise $m_3$, unless $SU(3)_c$ is embedded in a larger gauge group.
Note that the proton lifetime constraint (\ref{eq:const1}) implies that $\Lambda_{\rm pre}$ grows as $m_3^{1/2}$, while the scale where 
$\alpha_s$ loses perturbativity grows slightly faster than $m_3$. As a result, 
the proton is sufficiently long lived and $\alpha_s$  remains perturbative up to the compositeness scale if 
$m_3$ is  large enough.

Let us consider the more realistic Vquark mass spectrum discussed at the end of Section~\ref{sec:below}:
the weak-singlet down-type Vquarks have masses near 2 TeV, the other color-triplet Vquarks 
have masses    $m_{_{\cal Q}} \approx m_{_{7/6}} \approx m_{_{33}} \approx m_{_{4/3}}  \approx  m_{_{\cal U}}
 \approx 30$ TeV, and the color-octet and -sextet Vquarks have masses  
$m_{_{88}} \approx m_{_{6 \bar 6}} \approx  90$ TeV.
The proton decay constraint (\ref{eq:const1}) in this case is $\Lambda_{\rm pre}/ |C_8|^{1/4}   > 5 \times 10^4 \;  {\rm TeV}$.  
The 1-loop running gives $\alpha_s (m_{_{88}})  = 0.069$, and at a scale of $5 \times 10^4 \;  {\rm TeV}$
we find $\alpha_s \approx 0.44$, which is still perturbative given that the expansion parameter is $N_c \alpha_s/(4 \pi) \approx 0.11$. We estimate that the 2-loop running would increase $\alpha_s$, but $N_c \alpha_s/(4 \pi)$ would remain below 0.2. 

Thus,  for $|C_8|$ of order one, the perturbativity of $\alpha_s$ up to the compositeness scale, the mass spectrum of the composite vectorlike fermions, 
as well as the proton decay constraints allow $\Lambda_{\rm pre} \approx 5 \times 10^4 \;  {\rm TeV}$. 
The $\cal E$ and ${\cal L}_1$ Vleptons, and the charge $-1/3$ Vquarks have low enough masses ($m_{\cal E} \approx 260$ GeV, $m_{{\cal L}_1} \approx 1$ TeV, $m_{{\cal D}_{1,2}} \approx 2$ TeV) in this case to be probed in upcoming LHC searches. 
Further lowering $\Lambda_{\rm pre}$ can be consistent with the proton decay limit provided $m_{_{\cal Q}}$ is also lowered, but that would likely run into conflict with the LHC limit on the ${\cal L}_1$ Vlepton, whose mass is parametrically given by $m_{_{\cal Q}}^2  \left| y_{_{88}}^\prime  \,  \langle \phi_{_{88}} \rangle^*  + y_{_{6\bar 6} }^\prime \langle \phi_{_{6\bar 6} } \rangle^*  \right|^{-1} \! /N^2$, as shown in Section~\ref{sec:below}. 

If $|C_8|$  is as large as $10$, then proton decay sets a lower limit on $\Lambda_{\rm pre}$ of $9 \times 10^4  \;  {\rm TeV}$, which is still consistent with perturbativity since $\alpha_s \approx 0.4$ at that scale.
If $|C_8|$  is as small as $10^{-5}$, then the compositeness scale can be as low as $\Lambda_{\rm pre} \approx 3 \times 10^3  \;  {\rm TeV}$.
Note that once the heavy Vquark masses are fixed, the masses of the Vleptons and light Vquarks depend only logarithmically on $\Lambda_{\rm pre}$ through the Yukawa couplings in Eqs.~(\ref{eq:y0}) and (\ref{eq:yplog}), and their estimates are self-consistent as long as $ m_{_{88}} \ll \Lambda_{\rm pre}$.

\medskip  %\medskip\medskip

%%%%%%%%%%%%%%%%%%%%%%%%%%%%%%%%%%%%%%%%%%%%%%%%%%%%%%%%
%%%%%%%%%%%%%%%%%%%%%%%%%%%%%%%%%%%%%%%%%%%%%%%
\section{Novel signatures of baryon-number violation}
\label{sec:novel}\setcounter{equation}{0}

The prebaryon spectrum includes six ``right-handed neutrinos". Some of these neutral fermions may be lighter than the proton, so that decays of the type
\be
p \to \bar N^0 \pi^+      \;\;  ,   \;\;   p \to  \bar N^0 K^+   \;\;  ,   \;\;   n  \to  \bar N^0 \, \pi^0
\ee
are possible. The operator responsible for the first of these decays is 
\be
\frac{\widetilde C_N}{2 \Lambda_{\rm pre}^{8}}
\left( \Omega_{Q4}  \,  \Omega_{DL}  \right) 
\left(  \overline  \Omega_{QQ}^{(\overline  3,3)}  \,  \overline  \Omega_{QL}^{(3,3)}   \right)   
\left(  \Omega_{Q1} \Omega_{Q1}   \right)  
\left( \overline  \Omega_{D1}  \overline  \Omega_{41}  \right)       ~~,
\label{eq:pdecN}          
\ee
where $  \overline \Omega_{41}  \equiv N_R$, and we informally refer to the $N_R$ field as the $N^0$ particle.
Following the procedure presented after (\ref{eq:QQQLnoGUT}), we find that (\ref{eq:pdecN}) induces the proton-decay operator
\be
\frac{1}{M_{qqdN}^2}   \left(  u_L  d_L  \right)   \left(  d_R  \, N_R  \right)   ~~,
\label{eq:uLdLdRNR}
\ee
with a mass scale 
\be
M_{qqdN} = \frac{ 16 \pi^2  N^{3/2} \, N_c \, \Lambda_{\rm pre}^2}{  c_\phi \,  ( C_N \,  m_{_{\cal Q}} m_{_{33}} )^{1/2}}    ~~.
\ee
Here $C_N = N^3  N_c^2 \widetilde C_N$, because the counting of factors of $N$ and $N_c$ is the same as in the process discussed above (\ref{eq:Nfactors}).
If the $N^0$ mass, $m_N$, is below about 0.8 GeV, then operator (\ref{eq:uLdLdRNR}) induces the $p  \to   \bar N^0 \, \pi^+$ decay. The proton decay width, computed in general in \cite{Yoo:2021gql}, is given in our case by
\be
 \hspace*{-0.3cm} \Gamma \! \left( p \to  \bar N^0 \pi^+ \right)  = \frac{|\vec{q}_N| \, E_N \, W_0^2 (m_N) }{8\pi m_p\, |M_{qqdN}|^4} 
\left( 1- \frac{2 m_N^2 W_1 (m_N)}{m_p E_N W_0 (m_N)  }  +  \frac{m_N^2 W_1^2 (m_N)}{m_p^2 \, W_0^2 (m_N)}  \right) ,
\label{eq:piNwidth}
\ee
where $|\vec{q}_N|$ is the magnitude of the 3-momentum carried by $N^0$,
\be
|\vec{q}_N| \, = \frac{m_p}{2}
 \left( 1 -  \frac{  (m_N + m_{\pi^+})^2 }{m_p^2}  \right)^{\! 1/2}  \!  \left( 1 -  \frac{ (m_N - m_{\pi^+})^2 }{m_p^2}  \right)^{\! 1/2}  ~~,
 \label{eq:Nmomentum}
\ee
and the energy of $N^0$ is 
\be
E_N = \frac{m_p}{2} \left( 1 \! +  \frac{m_N^2 - m_{\pi^+}^2 }{m_p^2} \! \right)   ~~.
\ee
The hadronic matrix element for the proton decay induced by operator  (\ref{eq:uLdLdRNR}) can be written as 
\be
\langle \pi^+ |   \left(  u_L  d_L  \right) d_R \, | p \rangle_{q^2 = m_N^2} =  P_R \left( W_0 (m_N)  
- W_1  (m_N) \, q_\mu \gamma^\mu  \rule{0mm}{3.9mm} \right)  \! {\textsl u}_p   ~~,
 \label{eq:W0N}
\ee
where ${\textsl u}_p$ is the on-shell Dirac spinor of the proton.

There are some important differences between the $p \to \bar N^0 \pi^+ $ and $p \to e^+ \pi^0$  
widths displayed in (\ref{eq:piNwidth}) and (\ref{eq:pwidth}).  The $\bar N^0 \pi^+$ final state is phase-space 
suppressed due to the  $N^0$ mass, but it is enhanced by a factor of 2 due to the 
matrix elements for charged versus neutral pions. Furthermore, the form factors $W_0$ and $W_1$ evaluated at 
the lepton mass $m_N$ are likely to be larger than at $m_e$, because the probability for a quark and an 
antiquark to bind into a pion increases when their 3-momenta are smaller. 

The form factors have been computed on the lattice  \cite{Yoo:2021gql} only for lepton masses $m_\mu \approx 106$ MeV and 
$m = 0$, the latter being a good approximation for the positron mass compared to the proton mass. Their values at a scale of 2 GeV are $W_0 (m_N)   \approx -  (0.153 \pm  0.031) \,\,  {\rm GeV}^2$ and 
$W_1 (m_N)   \approx - (0.136 \pm  0.019)  \,\,  {\rm GeV}^2$ for $m_N = m_\mu$, while the values at $m_N = 0$ are slightly smaller in magnitude,
$W_0 (0)   \approx - (0.151 \pm  0.030) \,\,  {\rm GeV}^2$ and $W_1 (0)   \approx - (0.134 \pm  0.018) \,\,  {\rm GeV}^2$. Here for simplicity we added in quadrature the statistical and systematical errors.

For the ratio  $r(m_N) = \Gamma \! \left( p \to \bar N^0  \pi^+ \right)/ \Gamma \! \left( p \to e^+  \pi^0  \right)$ evaluated at $m_N = m_\mu$ for $M_{qqdN} = M_{qque}$ we find $r(m_\mu) = 1.93$.  As we do not know the values of the form factors for larger $m_N$, we cannot make precise statements about  $r(m_N)$ for $m_N$ closer to its upper limit of $m_p - m_{\pi^+}$. However, using the fact that $r(m_N) $ for $m_N > m_\mu$ is larger than the value obtained with form factors evaluated at the muon mass, we find $r(410 \; {\rm MeV} ) > 1$ and   $r(540 \; {\rm MeV} ) > 1/2$. Thus, $\bar N^0 \pi^+ $ is the dominant proton decay mode for a large range of $m_N$.

If $N^0$ is stable on the detector length scale, then the signal of $p \to \bar N^0 \pi^+$ is a single  $\pi^+$ of momentum $|\vec{q}_N|$, which is  given in (\ref{eq:Nmomentum}) as a function of the $N^0$ mass. A search for that signal in the case $m_N = 0$ has been carried out by the 
 Super-Kamiokande collaboration \cite{Super-Kamiokande:2013rwg} using a 22.5 kton water Cerenkov detector. The 90\% CL limit set by that search is $\tau (p \to \pi^+ \, \bar  \nu ) > 3.9\times 10^{32}$ years. A small excess of events has been observed in that search,
consistent with a $p \to \bar  N^0 \pi^+$ signal for $m_N$ in the 100 -- 200 MeV range. The full Super-Kamiokande data set, 
analyzed for example in the $p \to e^+ \pi^0$ channel \cite{Super-Kamiokande:2020wjk}, is 2.6 times larger than the one included in the existing $p \to \pi^+ \bar \nu $ search. Thus, an update of the search for $p \to  \pi^+ \bar \nu$ modified to be sensitive to  the $p \to \bar  N^0 \pi^+$ decay with $m_N > 100$ MeV could in principle lead to evidence for a stable $N^0$ and for baryon number violation. 

%%%%%%%%%%%
 \begin{figure}[b!]
  \begin{center}   \vspace*{0.2cm}
  %  \hspace*{.4cm}  
 \includegraphics[width=0.83\textwidth, angle=0]{ 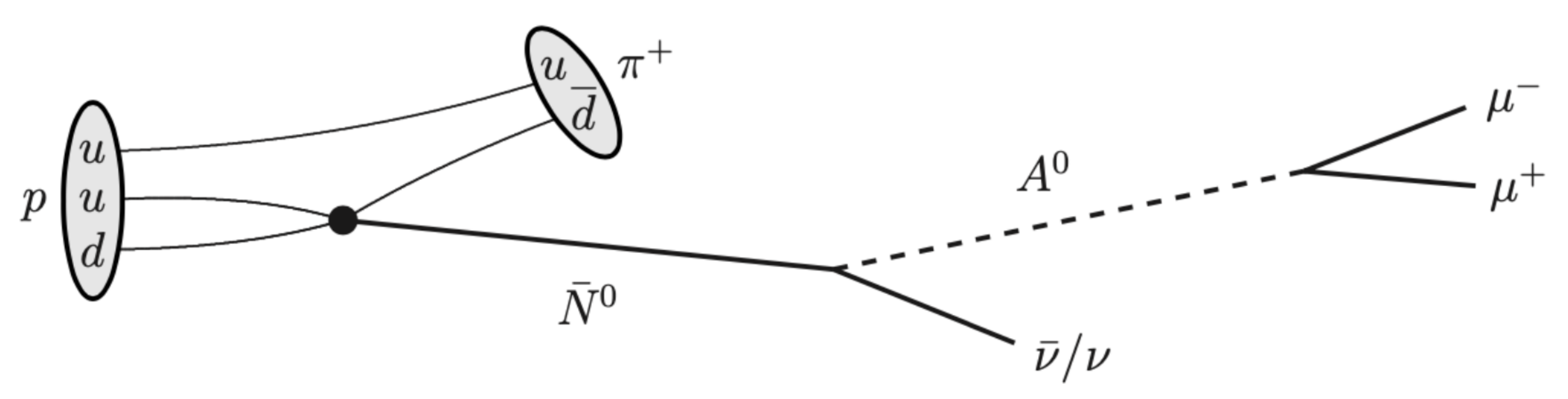}   
\vspace*{-0.2cm}
   \caption{Exotic proton decay $p \to \bar N^0 \pi^+$  followed by the decay of the antiparticle $\bar N^0$ of the ``right-handed neutrino" via a light pseudoscalar $A^0$. }
  \label{fig:ProtonDecayExotic}
  \end{center}
\end{figure}
%%%%%%%%%%%%

If $\bar N^0$ decays within the detector, various final states may be within the reach of existing or future proton decay experiments.
A distinct possibility is that  $\bar N^0$ is sufficiently longed-lived so that its decay happens several meters away from the location of the 
proton decay. In that case the signal is an isolated $\pi^+$ in conjunction with a $\bar  N^0$ decay in a different part of the detector. That decay is model dependent. An example is the process $\bar N^0 \to A^0 \bar \nu$ (and also $\bar N^0 \to A^0 \nu$ if $N^0$ has a Majorana mass), where  $A^0$ is a composite pseudo-Nambu-Goldstone boson, which is present in the model for quark and lepton compositeness (see Section 3) and could be lighter than $N^0$. 
The $A^0$ may subsequently decay  into a $\mu^+\mu^-$ pair (see Figure \ref{fig:ProtonDecayExotic}), potentially with a displaced vertex.

Three future experiments are currently under construction. Hyper-Kamiokande~\cite{Hyper-Kamiokande:2018ofw} is a planned
water Cerenkov detector like Super-Kamiokande, but about seven times larger. DUNE~\cite{DUNE:2015lol} is a liquid Argon
detector, and  JUNO~\cite{JUNO:2015zny} is a liquid scintillator detector. 
Due to its excellent particle identification capabilities, DUNE will be the most sensitive experiment to a signal like the one discussed above, consisting of an isolated $\pi^+$ and a displaced $\mu^+\mu^-$ pair. Measuring the invariant mass and momentum of the $\mu^+\mu^-$ pair will determine the mass and momentum of $A^0$.
Since the $\pi^+$ and $\bar N^0$ momenta are back-to-back, the direction of the $\bar N^0$ momentum is known, and requiring that it intersects the  direction of the $A^0$ momentum would reject various backgrounds. 
Furthermore, the angle between the $\bar N^0$ and $A^0$ momenta, combined with the $\pi^+$ momentum measurement,
can be used to determine $m_N$, and also to reconstruct the proton mass, which further reduces the backgrounds.
The $p \to \bar N^0 K^+$ decay followed by a displaced decay of  $\bar N^0$ is also interesting, and provides special opportunities at DUNE because the liquid Argon time-projection chambers allow clean identification of kaons.

%\bigskip

%%%%%%%%%%%%%%%%%%%%%%%%%%%%%%%%%%%%%%%%%%%%%%%%%%%%%
%%%%%%%%%%%%%%%%%%%%%%%%%%%%%%%%%%%%%%%%%%%%%%%%%%%%%
\section{Conclusions}
\label{sec:conc}

Uncovering deeper structure has been a hallmark of fundamental physics until the discovery of quarks. It is compelling to continue 
inquiring about substructure, especially one that would explain the existence of quarks and leptons in terms of some underlying degrees of freedom, conventionally called preons. The gauge representations of the SM imply that the interactions that hold the preons together must be chiral, which complicates their study given that there are no known methods for computing the behavior of strongly-coupled chiral gauge theories. 
Nevertheless, theories of quark and lepton compositeness can be constructed and analyzed at a qualitative level. In a model of this type based on a chiral $SU(15)$ gauge symmetry, proposed in \cite{Dobrescu:2021fny}, the quarks and leptons are prebaryons 
made of three preons, while the Higgs doublets are bound states of two prebaryons.

We have shown here that proton-decay operators are likely to be generated by the $SU(15)$ dynamics. Specifically, we identified the 8-prebaryon operators (\ref{eq:QQQLnoGUT}) and (\ref{eq:udQL}), which conserve each preon number, and thus are likely to be generated at the compositeness scale $\Lambda_{\rm pre}$.
These dimension-12 operators include pairs of prebaryons that form scalar bound states, thus generating at lower scales some gauge invariant operators involving three quarks, one lepton and two composite scalars. The latter have VEVs related to the 
mass $m_{_{\cal Q}}$  of composite Vquarks, and thus proton-decay operators are generated with a suppression scale that depends on $\Lambda_{\rm pre}$ and $m_{_{\cal Q}}$. The SuperKamiokande limit on the $p \to e^+ \pi^0$ lifetime implies
$\Lambda_{\rm pre} > 10^4 \, {\rm TeV}^{1/2} \, C_8^{1/4} \, m_{\cal Q}^{1/2}$, where $C_8$ is the unknown coefficient of the 8-prebaryon operator with the appropriate powers of $N=15$ and $N_c =3$ factored out.

The spectrum of prebaryons also includes Vleptons, whose masses arise from the same VEVs that generate the Vquark masses. We have shown that the lower limit of the Vlepton masses set by LHC searches implies $m_{_{\cal Q}} \gtrsim 30$ TeV,
which in turn implies $\Lambda_{\rm pre} > 5 \times 10^4 \, {\rm TeV}\, C_8^{1/4} $.
Thus, even if the unknown rescaled coefficient  is as small as $10^{-5}$, the compositeness scale is still severely constrained, $\Lambda_{\rm pre} \gtrsim 3000$ TeV. This raises the question of whether the $SU(15)$ dynamics can indeed generate the hierarchy between the weak scale, which is given by a 
di-prebaryon VEV, and the compositeness scale. Although we do not have the tools to answer that question, it is worth recalling that the  energy scales measured in deuteron systems are a few orders of magnitude lower than the QCD scale.
The above bound on $\Lambda_{\rm pre} $, set by proton lifetime limits, is about two orders of magnitude higher than the bounds derived from flavor-changing processes \cite{Chivukula:1997iw} in generic models of compositeness.

The chiral $SU(15)$ model predicts that there are nine composite Vquarks in $SU(3)_c$ representations which imply that QCD loses asymptotic freedom. We have shown, however, that the QCD coupling continues to be  perturbative up to the compositeness scale. At larger scales, the QCD running is no longer affected by the Vquarks, and instead is governed by the preon representations under $SU(3)_c$. We do not address here the physics above the compositeness scale, except to point out that $SU(3)_c$ can be embedded in a larger gauge group at scales above $10^8$ GeV, and thus its coupling may regain asymptotic freedom.

We have pointed out that novel proton decay modes, especially into a $\pi^+$ and the antiparticle of a heavy right-handed neutrino, $N^0$, are predicted to have a large branching fraction in this model if the mass of $N^0$ is below $\sim 0.5$ GeV.
An update of the Super-Kamiokande search for $p\rightarrow \pi^+ \bar \nu$ modified to be sensitive to the $p\rightarrow \bar N^0  \pi^+$ decay with $m_N>100\phantom{t}\rm{MeV}$ could in principle lead to evidence for a stable $N^0$.
More exotic signals may be produced if $\bar N^0$ decays inside the detector. For example, 
the decay $\bar N^0 \to A^0 \bar \nu/\nu $, where  $A^0$ is a pseudo-Nambu-Goldstone boson, followed by $A^0 \to \mu^+\mu^-$, would produce a spectacular signal in the planned DUNE detector, due to the particle identification capabilities of the liquid Argon time projection chambers.

Besides discovery modes in proton decay experiments, the model of composite quarks and leptons predicts vectorlike fermions that may be light enough to be discovered at the LHC. A weak-doublet Vlepton of mass above 1 TeV, and a weak-singlet Vlepton of mass as low as 260 GeV can be probed in the current Run 3. In addition, the LHC experiments will become sensitive to  two weak-singlet Vquarks of charge $-1/3$ with masses that can be as low as 2 TeV. 

\bigskip   %\bigskip    %\bigskip 
\smallskip

%%%%%%%%%%%%%%%%%%%%%%%%%%%%
\noindent
{\it  Acknowledgments:}  {\small  We would like to thank Bhupal Dev, Patrick Fox, Roni Harnik, Pedro Machado, Jennifer Raaf, and Mike Wagman for insightful comments. 
The graphics have been generated using Axodraw~2 \cite{Collins:2016aya}.
This work is supported by Fermi Research Alliance, LLC under Contract DE-AC02-07CH11359 with the U.S. Department of Energy.
}

%%%%%%%%%%%%%%%%%%%%%%%%%%%%%%%%%%%%%%%%%%%%%%%%%%%%%
%%%%%%%%%%%%%%%%%%%%%%%%%%%%%%%%%%%%%%%%%%%%%%%%%%%%%

%%%%%%%%%%%%%%%%%%%%%%%%
\bibliographystyle{JHEP}

\end{document}